\documentclass[twocolumn,5p]{elsarticle}

\usepackage{graphicx}
\usepackage{amsmath}
\usepackage{amssymb}

\def\pd{\partial}
\def\m{\mu}

\newcommand{\ie}{{\it i.e.}}

\newcommand{\be}{\begin{equation}}
\newcommand{\ee}{\end{equation}}
\newcommand{\br}{\begin{eqnarray}}
\newcommand{\bea}{\begin{eqnarray}}
\newcommand{\eea}{\end{eqnarray}}
\newcommand{\er}{\end{eqnarray}}
\newcommand{\ba}{\begin{array}}
\newcommand{\ea}{\end{array}}
\newcommand{\bi}{\begin{itemize}}
\newcommand{\ei}{\end{itemize}}
\newcommand{\bn}{\begin{enumerate}}
\newcommand{\en}{\end{enumerate}}
\newcommand{\bc}{\begin{center}}
\newcommand{\ec}{\end{center}}

\newcommand{\Eq}[1]{Eq.~(\ref{#1})}

\newcommand{\beq}{\begin{equation}}
\newcommand{\eeq}{\end{equation}}

\newcommand{\gsim}{\lower.7ex\hbox{$\;\stackrel{\textstyle>}{\sim}\;$}}
\newcommand{\lsim}{\lower.7ex\hbox{$\;\stackrel{\textstyle<}{\sim}\;$}}

\def\mysection#1{{\bf #1.} }

\begin{document}

\title{Dark Supersymmetry}

\author[NICPB]{Matti Heikinheimo}
\address[NICPB]{National Institute of Chemical Physics and Biophysics, Ravala 10, 10143 Tallinn, Estonia}

\author[NICPB]{Antonio Racioppi}

\author[NICPB,Tartu]{Martti Raidal}
\address[Tartu]{Institute of Physics, University of Tartu, Estonia}

\author[NICPB]{Christian Spethmann}

\author[Jyv,Hel,UK]{Kimmo Tuominen}
\address[Jyv]{Department of Physics, Jyv\"askyl\"a University, P.O.Box 35, FIN-40014 Jyv\"askyl\"a, Finland}
\address[Hel]{Helsinki Institute of Physics, P.O.Box 64, FIN-00014, Helsinki University, Finland}
\address[UK]{Department of Physics and Astronomy, University of Southampton, Southampton, SO17 1BJ, UK }

\date{\today}

\begin{abstract}
We propose a model of Dark Supersymmetry, where a supersymmetric dark sector is coupled to
the classically scale invariant non-supersymmetric Standard
Model through the Higgs portal. The dark sector contains a mass scale that is protected against
radiative corrections by supersymmetry, and the portal coupling mediates this scale to the Standard
Model, resulting in a vacuum expectation value for the Higgs field and the usual electroweak
symmetry breaking mechanism. The supersymmetric dark sector contains dark matter
candidates, and we show that the observed dark matter abundance is generated for a natural choice of
parameters, while avoiding the current experimental bounds on direct detection. Future experiments can
probe this scenario if the dark sector mass scale is not too high.
\end{abstract}


\maketitle

\section{Introduction}
The discovery~\cite{Aad:2012tfa} of the Higgs boson~\cite{Englert:1964et} at the LHC
proves experimentally that the origin of electroweak (EW) symmetry breaking is Standard
Model (SM)-like within ${\cal O}(20\%)$ accuracy~\cite{Giardino:2013bma}.
The main questions to answer now are what generates the electroweak scale, and what stabilizes it against radiative corrections?

Historically, the most popular idea to explain the naturalness of the EW scale has been
the extension of space-time symmetries with supersymmetry~\cite{Nilles:1983ge}.
If supersymmetry is exact, the non-renormalization theorem guarantees that this scale
remains stable also at the quantum level, independent of its origin in the superpotential.
Soft supersymmetry breaking does not change this result, as long as the soft terms are comparable
to the EW scale. Unfortunately, the LHC has not detected any signal of the popular models of
supersymmetry, rendering those solutions to the naturalness of EW scale unnatural~\cite{Strumia:2011dv}.
This led us to argue~\cite{Heikinheimo:2013fta} that the LHC results may imply a paradigm change
in searches for new physics, which we know must exist due to the existence of dark matter
(DM)~\cite{Ade:2013lta}.

Indeed, the SM itself is a physically natural theory~\cite{Heikinheimo:2013fta,Farina:2013mla,Dubovsky:2013ira}
as long as it contains only renormalizable interactions, \ie, large higher-dimensional
operators are absent, and the new physics scale it couples to does not exceed the EW scale
by more than 1-2 loop factors. The experimental results obtained during the last decades
(flavor physics, precision electroweak experiments, non-observation of proton decay)
support the absence of higher dimension operators with remarkable precision.
Additionally, the absence of a large contribution to the Higgs mass in the Standard Model
Lagrangian has now also become a striking observation. The traditional view on these
results is that they are in conflict, and much effort has been expended to reconcile
them with each other.

In this paper we take these results as a suggestion that the underlying theory of
nature might be classically scale invariant~\cite{Coleman:1973jx}.
If classical scale invariance is taken to be an exact symmetry of the underlying
Lagrangian, all explicit mass scales must be absent at the fundamental level.
The scale invariance of the SM is broken by quantum effects, resulting in the running
of the coupling constants and in the appearance of the
QCD confinement scale and baryon masses. Consequently, any other distinct scale such as
the EW scale, must then have a dynamical origin. In this setting the
Higgs mass and vacuum expectation value, and therefore all the masses of the elementary
particles of the SM, must be generated by a dynamical mechanism.

There are only two known generic mechanisms for dynamical generation of the EW
scale, based on weak and strong dynamics.  The first class is based on
dimensional transmutation {\it a la} the Coleman-Weinberg mechanism~\cite{Coleman:1973jx},
and the second is based on strong dynamics of asymptotically free gauge theories {\it a la}
QCD, as for example in Technicolor~\cite{Weinberg:1979bn}. Both weak and strong
dynamical EW symmetry breaking mechanisms have been extensively studied in last decades as
alternatives to the Higgs mechanism in the SM, and found to be strongly constrained.

After the first phase of the LHC, the resulting possible picture of particle physics is the following:
There exist two sectors, the classically conformal SM sector and the dark sector. The latter
is manifested by the existence of DM, and consists of SM singlets but may have complicated
internal dynamics. The two sectors are connected only via the Higgs portal,
\be
\lambda_{HS} |H|^2 |S|^2,
\label{portal}
\ee
where $S$ is a scalar mediator field that couples to the dark sector fields\footnote{The physics
content and predictions of our scenario are very different from the simplest models where
$S$ itself is a scalar DM particle due to imposing an additional $Z_2$~\cite{Silveira:1985rk} or
$Z_3$~\cite{Belanger:2012zr} symmetry.}.
The internal structure of the dark sector is such that it generates dynamically a physical scale
and a singlet vacuum expectation value (VEV) $\langle S \rangle$, that induces the negative
SM Higgs boson mass term  $-\mu^2 |H|^2$ via \Eq{portal} with negative $\lambda_{HS}$.
It is worth noting that adding the singlet $S$ {\it improves} the vacuum stability of the SM Higgs potential
due to the renormalisation effects of the interaction in \Eq{portal}~\cite{Kadastik:2011aa}.
Natural mechanisms for generating the scale $\langle S \rangle \sim {\cal O}(1)$~TeV are
Dark Technicolor~\cite{Heikinheimo:2013fta,Hur:2011sv} and a Dark Coleman-Weinberg
mechanism~\cite{Hempfling:1996ht}.
While the simplest Coleman-Weinberg or Technicolor models are not viable alternatives to
the SM Higgs mechanism, they are both perfectly natural generic candidates for scale generation
in the dark sector, where stringent phenomenological constraints do not arise. The generic predictions of
this setup are small mixing of the SM Higgs boson with the singlet scalar, which could be tested
in Higgs factories such as the ILC or CLIC, and suppressed DM coupling to baryonic matter.

Models of this type, consisting of the classically scale invariant SM sector without the Higgs
mass term, and a dark sector that dynamically generates a scale and transmits it to the SM via
the Higgs portal, allow for a non-conventional way to address the hierarchy problem. In this setting all mass scales are generated dynamically, and protected from radiative corrections by classical scale invariance. It has to be recognized, however, that classical scale invariance can only protect the largest physical mass scale in the full theory, since that scale sets the amount of breaking of this symmetry. If the EW scale is not the largest scale associated with massive particles, then there must be additional mechanisms in play to ensure the naturalness of the EW scale by cancelling the
radiative corrections to the Higgs mass from the new physics. Clearly this is in contrast with {\it e.g.} GUT models, where a large number of heavy particles exist, so to address the hierarchy problem in this manner one has to give up the thought of embedding the SM in a unified gauge theory at some UV scale. Consequently, also the Planck scale can not be associated with massive particles at $m_{\rm Planck}$ in this case, but instead must be understood as a scale where a different theory takes over the SM.
This theory could be a non-local theory as suggested in~\cite{Dubovsky:2013ira}. The possibility to include heavy Majorana neutrinos, as required in see-saw models, while preserving the physical naturalness of the SM has been discussed in \cite{Farina:2013mla}.

In this work we propose a third generic alternative for natural generation of the EW scale
in the dark sector, Dark Supersymmetry. We assume that an initially supersymmetric theory
is broken at some high scale in such a way that the visible sector with the SM particle
content and symmetries is not supersymmetric at all.
An analogous setup can, for example, be realized if the SM particles are confined to a
non-supersymmetric brane (due to complicated dynamics)
~\cite{Cicoli:2011yy} or in $B-L$ extensions of the MSSM \cite{Arnold:2012wm}.
The dark sector on the other hand retains its supersymmetry. The most minimal model, which we call Dark Supersymmetry, is
obtained by assuming the dark sector to consist of only one singlet chiral superfield $S$.
The mass scale in the dark sector is then given by the mass parameters in the superpotential
of the singlet, since its scalar component $s$ naturally obtains a VEV of the order of that scale
without breaking supersymmetry. Whatever the scale is, its stability is guaranteed by supersymmetry.
As described previously, this scale triggers EW symmetry breaking in the SM via the
portal interaction in \Eq{portal} and, as a back-reaction, breaks supersymmetry in the dark sector explicitly\footnote{Here we imagine that SUSY in the dark sector is only broken as a back-reaction by the portal coupling. If there is additional SUSY breaking in the dark sector itself, this breaking could potentially destabilize the dark sector mass scale. Therefore such breaking, if it is present, should not happen at a scale too high above the electroweak scale.}.

We work out the phenomenology of this model. In particular, we find the parameter space of
Dark Supersymmetry that is consistent with SM Higgs phenomenology and with the DM relic
abundance of the Universe. Based on those results, we discuss implications of Dark
Supersymmetry on Higgs boson phenomenology and on DM direct detection.

The structure of this paper is the following: In the next section (II) we formulate Dark
Supersymmetry and work out the mass spectrum and couplings. In section III we study
the phenomenology of the model. We discuss our results and conclude in section IV.

\section{Dark Supersymmetry}

We start by studying the properties of Dark Supersymmetry in the limit of unbroken supersymmetry.
Many of our main results concerning scales and EW symmetry breaking are explicit already in this case.
Since the Higgs portal,  \Eq{portal}, breaks supersymmetry, this limit is strictly speaking unphysical.
Therefore, we work out the properties of most general scalar potential in a model with softly
broken Dark Supersymmetry in the next subsection.

\subsection{Description of supersymmetric dark sector}

We assume the dark sector to consist of one singlet chiral superfield $S$ with renormalizable superpotential
\be
 W = a_S S + \m_S S^2 + \lambda_S S^3.
 \label{W}
\ee
In principle the dark world can be much more complicated containing many superfields, which also carry internal
quantum numbers. In this case the term $a_S S$ must be forbidden and the quadratic and cubic terms must contain
different superfields, but this will not affect our main result of transmitting the SUSY scale from $W$ to the SM.
We therefore work with the most minimal superpotential and comment on the absence of additional terms if necessary.

The scalar potential corresponding to \Eq{W} is given by
\bea
 V_s &=& f_S^* f_S = \left| \pd_s W(s) \right|^2 \nonumber \\
   &=& \left| a_S  + 2 \m_S s + 3 \lambda_S s^2 \right|^2,
\eea
where we assume for simplicity that all the parameters $a_S$, $\mu_S$ and $\lambda_S$ are real.
The minimum is given by the solutions of
\be
 f_S^*= f_S=0.
\ee
We parametrize $s=\frac{s_R + i s_I}{\sqrt 2}.$ Without loss of generality we can assume that only $s_R$ gets
a vacuum expectation value (VEV),
\be
 v_{\text{susy} 1,2} = -\frac{\sqrt{2} \mu_S}{3 \lambda_{S}}
                      \pm \frac{\sqrt{2} \sqrt{\mu_S^2-3 a_S \lambda_{S}}}{3 \lambda_S},
\label{vsusy12}
\ee
where we assume $\mu_S^2-3 a_S \lambda_{S} >0$. The scalar potential $V_s$ is symmetric under
\be
 s \to -s - \frac{2 \mu_S}{3 \lambda_{S}} ,\label{symmofV}
\ee
which is the axial symmetry around the vertical axis of the local maximum
\be
 s_{R\text{max}} =-\frac{\sqrt{2} \mu_S}{3 \lambda_{S}}.
\ee
The VEVs in \Eq{vsusy12} are obviously breaking the axial symmetry.

The physical particles of the model are the singlino, $\psi_S$, and the mass-degenerate scalars, $s_R$ and $s_I.$
The mass spectrum is given by
\bea
 m_s &=& 2 \sqrt{\mu_S^2-3 a_S \lambda_{S}}, \\
 m_{\psi_S} &=& m_s  .
\label{susymasses}
\eea
Notice that even if $\langle s \rangle \neq 0$, SUSY is not broken because $\langle f_S \rangle =0$.
This is analogous to the NMSSM where the singlet also gets a VEV.

Let us comment on some special cases.
If $\mu_S=0$, the scalar potential will be
\be
  \left. V_s \right |_{\mu_S=0} = \left | a_S  + 3 \lambda_S s^2 \right |^2,
\ee
which still has two minima different from zero as long as $a_S <0$.
More interesting is the case $a_S=0$. In such a case the two equivalent VEVs are
\bea
 v_{\text{susy }1} &=& 0, \\
 v_{\text{susy }2} &=& -\frac{2\sqrt{2} \mu_S}{3 \lambda_{S}}.
\eea
So we get a VEV equal to zero and one different from zero, without using a negative mass
squared parameter; the origin of the symmetry breaking now lies in the $s^3$ terms of $V_s$.
Notice that even if  $v_{\text{susy }1}=0$, the symmetry of the scalar
potential under the transformation in \Eq{symmofV}  is broken.
This will play an important role in the next section.

\subsection{The full scalar potential}

In the previous subsection, we studied the properties of the supersymmetric dark
sector in isolation without taking into account the Higgs portal.
Now we consider the full scalar potential of the SM and the dark sector, also including the Higgs terms.
It is given by
\be
 V =  V_s  + V_H + V_{sH} + V_\text{soft}, \label{V}
\ee
where
\bea
 V_H    &=& \lambda | H |^4, \\
 V_{sH} &=& \lambda_{sH} | s |^2 | H |^2,  \label{portal2} \\
 V_\text{soft} &=& b_1 s^\dagger s + (k s + b_2 s^2 + a_\lambda s^3 + h.c.) .
\eea
Here $V_H$ is the scale invariant Higgs potential of the SM with $\lambda>0$ and $V_{sH}$
is the Higgs portal connecting the two sectors of the model.
For simplicity we neglect the potentially allowed term $(s^2+s^{\dagger 2})|H|^2$
that can be forbidden by internal $S$ symmetries. This does not affect our results.
Since SUSY is only a symmetry of the dark sector, it is not necessary to
introduce SUSY breaking terms $V_\text{soft}$. However, since such terms could be induced
by the same physics generating the portal $V_{sH}$ or by some other interaction of
the dark sector, we include $V_\text{soft}$ in order to the keep the model minimal
and at the same time as general as possible.

Because of $V_{sH}$ and $V_\text{soft}$, the scalar potential is not symmetric under
the transformation in \Eq{symmofV}
anymore, but only under the gauge symmetries involving the Higgs.
This means that we get only one absolute minimum in the singlet sector, and no longer a pair of equivalent minima.
In order to achieve EW symmetry breaking we require that the effective Higgs square mass
term---induced by $V_{sH}$ via the singlet VEV---is negative.
So we have to assume ${\lambda_{sH}<0}$ and the following condition,
\be
 \lambda_{sH}>-\lambda -9 \lambda_{S}^2,
 \label{quarticpositive}
\ee
in order to have the quartic term always positive. However this condition is not enough:
since we have $s^3$ terms we need also to ensure that the are no directions in which the
quartic term is zero. This is done by imposing
\be
 \lambda_{sH}^2<36  \lambda \lambda_{S}^2,
\ee
which is a stronger condition than the one in  \Eq{quarticpositive}.

As usual, we have $\langle G^0\rangle, \langle G^\pm\rangle=0$. Considering again only real parameters
in \Eq{V}, we can still assume $\langle \Im(s)\rangle=0$.
So the part of the scalar potential relevant for the minimization is
\bea
&&\hspace{-1.3cm}
V_{s_R,\phi}  = a_S^2 +\sqrt{2} {\bar k} s_R + \left( {\bar b} + \frac{{\bar \mu_S}^2}{2} \right) s_R^2 + \\
        &&\hspace{-1cm}
 \left( \frac{a_\lambda}{\sqrt{2}}+3 \sqrt{2} \lambda_S \mu_S \right) s_R^3 + \frac{9 \lambda_S^2s_R^4}{4}+
              \frac{\lambda \phi^4}{4}+\frac{1}{4} \lambda_{sH} \phi^2 s_R^2 ,\nonumber
\eea
where
\bea
 {\bar b} &=& 3 a_S \lambda_S+b_2 ,\\
 {\bar k} &=& 2 a_S \mu_S+k ,\\
 {\bar \mu_S}^2&=& b_1+4 \mu_S^2.
\eea
The corresponding minimization equations read
\bea
&&
\lambda  v^2 +\frac{\lambda_{sH} v_s^2}{2} = 0,\\
&&v_s^3 \left(9 \lambda_S^2-\frac{\lambda_{sH}^2}{4 \lambda }\right) + v_s^2 \left(\frac{3a_\lambda }{\sqrt{2}}+9 \sqrt{2} \lambda_S \mu_S \right)+\nonumber\\
&&v_s (2 {\bar b}+{\bar \mu_S}^2)+ \sqrt{2} {\bar k} = 0,
\eea
where
\be
 \langle\phi\rangle=v/\sqrt2, \qquad \langle s\rangle=v_s/\sqrt2=\langle s_R\rangle/\sqrt2.
\ee
The extremization equations and the diagonalization of the square mass matrix, $M^2_{s_R \phi}$, can be solved exactly.
In order to ensure that they are the minima, the eigenvalues of $M^2_{s_R \phi}$ must be all positive.
However we know that the mixing angle $\theta_{sH}$ between the singlet and the Higgs,
\be
 \theta_{sH} = \frac{1}{2} \arctan\left[\frac{2 (M^2_{s_R \phi})_{12}}{(M^2_{s_R \phi})_{11}-(M^2_{s_R \phi})_{22}}\right],
\ee
is usually small. So we can threat $\lambda_{sH}$ as a small parameter and give results
as expansions in $\lambda_{sH}$ up to the second order. So we get the following VEVs:
\bea
 v^2 &\simeq& -\frac{\lambda_{sH} v_{R} ^2}{2 \lambda }, \\
 v_s &\simeq& v_{R}+ \frac{\lambda_{sH}^2  v_{R}^3}{4 \lambda    m_{s_R}^2},
\eea
where $v_R$ is the singlet VEV computed in absence of the Higgs portal, given by
\be
9 \lambda_S^2 v_R^3 +  \left(\frac{3 a_\lambda }{\sqrt{2}}+9 \sqrt{2} \lambda_S \mu_S \right) v_R^2 +(2 {\bar b}+{\bar \mu_S}^2)v_R +\sqrt{2} {\bar k} = 0.
\ee
The square mass matrix becomes
\bea
 M^2_{s_R \phi} \simeq \left(
\begin{array}{cc}
 m_{s_R}^2 + \lambda_{sH}^2 \Delta_{s_R}&
 -v_R^2 \lambda_{sH} \sqrt{\frac{|\lambda_{sH}|}{2\lambda }} \\
 -v_R^2 \lambda_{sH} \sqrt{\frac{|\lambda_{sH}|}{2\lambda }}
 & -v_R^2 \lambda_{sH}
\end{array}
\right),
\eea
where
\bea
&&\hspace{-1cm} m_{s_R}^2 = \frac{3}{2} v_R \left(\sqrt{2} a_\lambda+6\lambda_S\left(\sqrt{2}
   \mu_S+2\lambda_Sv_R\right)\right)-\frac{\sqrt{2}
   {\bar k}}{v_R}, \nonumber\\
&&\hspace{-1cm}  \Delta_{s_R}=\frac{ v_R \left(2 \sqrt{2} {\bar k} +  m_{s_R}^2 v_R +18 \lambda_S^2 v_R^3\right)}{4 \lambda   m_{s_R}^2}.
\eea
The Higgs boson and the real singlet mass eigenvalues are
\bea
&& \hspace{-1.1cm} m_h^2 \simeq |\lambda_{sH}| \, v_{R} ^2, \\
&& \hspace{-1.1cm} m_s^2 \simeq m_{s_R}^2+\frac{\lambda_{sH}^2 \left(2 v_R \left(\sqrt{2} {\bar k}+9
   \lambda_S^2 v_R^3\right)+m_{s_R}^2 v_R^2\right)}{4
   \lambda  m_{s_R}^2},
    \label{msphys}
\eea
and their mixing angle for the corresponding eigenstates is given by
\be
 \theta_{sH} \simeq \frac{\sqrt{|\lambda_{sH}|} m_h^2}{ \sqrt{2} \sqrt{\lambda } m_{s_R}^2}.
 \label{mixing}
\ee
Notice that the mixing angle has a double suppression by the smallness of $\sqrt{|\lambda_{sH}|}$ and by $m_h^2/m_{s_R}^2.$
The mass for the imaginary part of the singlet is given by
\be
 m_A^2 \simeq m_{s_I}^2+
  \frac{\lambda_{sH}^2}{4 \lambda m_{s_R}^2}\left[\sqrt{2} {\bar k} v_R-\left(\frac{9 a_\lambda}{\sqrt{2}}+3 \sqrt{2} \lambda_S \mu_S\right) v_R^3\right]     \label{mA},
\ee
where
\be
 m_{s_I}^2 = 2 \left(6\lambda_S v_R \left(2 \sqrt{2} \mu_S+3 \lambda_S v_R\right)+{\bar \mu_S} ^2\right)-m_{s_R}^2,
\ee
is the mass of the CP odd scalar in absence of the Higgs portal.
Notice that there is an accidental $Z_2$ symmetry in the singlet pseudoscalar sector coming from $s\leftrightarrow s^\dagger.$
This is nothing but the $CP$ symmetry. Because of this, the singlet pseudoscalar (now labelled as $A$) does
not mix with the Higgs Goldstone bosons. Due to this accidental symmetry, $A$ also becomes a viable
DM candidate, in addition to $\psi_S.$

The mass of the singlino $\psi_S$ will be changed as well since the value of the singlet VEV is changed
\be
 m_{\psi_S} = \left| -2 \mu_S-3 \sqrt{2}\lambda_S \left(v_R + \frac{\lambda_{sH}^2 v_R^3}{4 \lambda  m_{s_R}^2} \right) \right| \label{mpsi}.
\ee

Finally, we comment on the special case $a_S=0$ and $V_\text{soft}=0$.
Because of the Higgs portal, \Eq{portal2}, we now have only one global minimum of the scalar
potential that is always different from zero. This is because the Higgs portal breaks SUSY.
It is interesting to notice that the VEVs of $s$ and $H$ will be both different from zero
without using any negative mass squared term.

\section{Phenomenology of Dark Supersymmetry}

The Dark Supersymmetry framework described in the previous subsections reproduces the
effective negative Higgs mass term that induces spontaneous  EW symmetry breaking
exactly as in the SM. It appears because the Higgs boson mixes with the dark singlet.
The mixing angle given by \Eq{mixing} is naturally of the order ${\cal O} (10^{-3})$
if the SUSY (and the DM) scale is of the order ${\cal O}$(TeV). This small mixing is unobservable at
the LHC. The ILC may observe it provided the SUSY mass scale is not too far from the EW scale.

 Accidentally, the model contains two viable DM candidates: The singlino $\psi_S$ and the singlet pseudoscalar $A.$
 While the former is protected by supersymmetry, the latter is accidentally stable due to $CP$ symmetry.
 Unless  $2m_{\psi_S}<m_A$ so that the pseudoscalar decays to two $\psi_S,$ the pseudoscalar is also stable.

This accidental feature makes the DM phenomenology of the model unexpected and interesting. The singlino couples to the SM matter only via the
Higgs-singlet mixing that makes the singlino component of the DM undetectable in the recent direct detection experiments. The spin-independent nucleon-singlino scattering cross section is
\be
 \sigma_\text{SI}^\psi \simeq \frac{g_s^2 \sin^2(2 \theta_{sH})}{8\pi} \frac{m^4_N f^2}{v^2} \left(\frac{1}{m_h^2}-\frac{1}{m_s^2}\right)^2,
\label{sigmaSIpsi}
\ee
where $g_s= - 3\sqrt2\lambda_S$ is the Yukawa coupling between
$s$ and $\psi_S$, $m_N$ is the nucleon mass and $f$ is the nucleon
matrix element. However, the pseudoscalar component of DM couples
to the SM Higgs directly via the portal \Eq{portal2} that is suppressed
only by the smallness of the coupling ${|\lambda_{sH}|}$, and results in a scattering cross section
\be
 \sigma_\text{SI}^A \simeq \frac{\lambda_{sH}^2}{4\pi} \frac{m^4_N f^2}{m_A^2 m_h^4}.
\ee
This makes the model testable.

\begin{figure}[t]
\begin{center}
\includegraphics[width=0.5\textwidth]{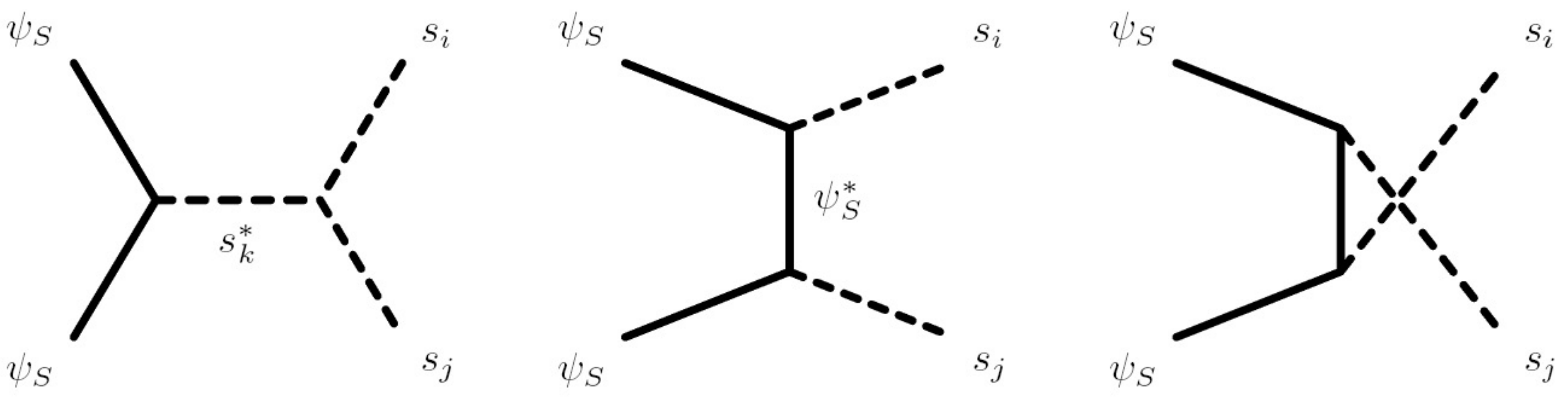}
\caption{$\psi_S$ dominant annihilation modes. Since $\psi_S$ is a Majorana spinor there are no fermion flow arrows in the diagrams.}
\label{fig:psiannihilation}
\end{center}
\end{figure}

\begin{figure}[t]
\begin{center}
\includegraphics[width=0.5\textwidth]{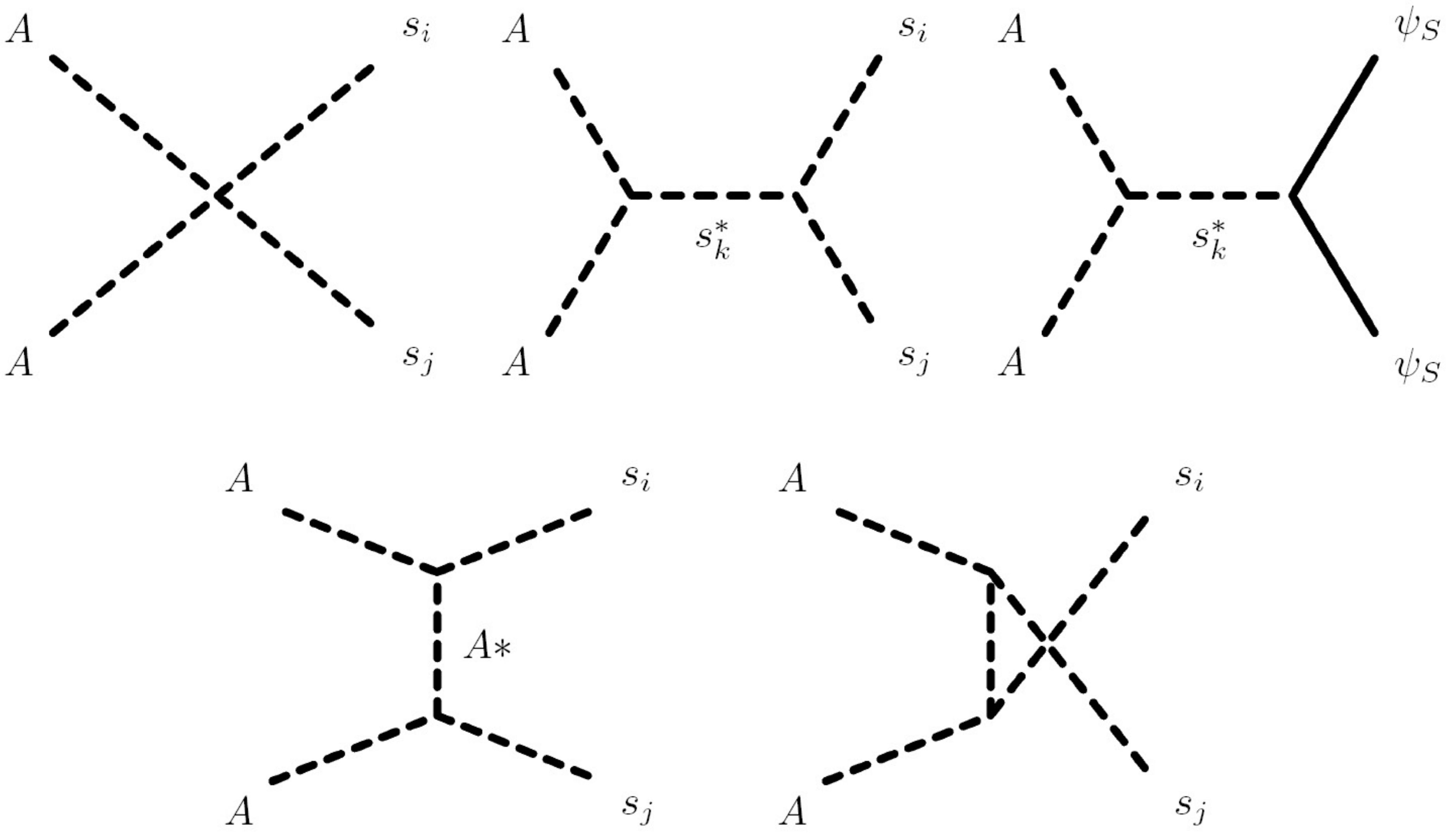}
\caption{$A$ dominant annihilation modes.}
\label{fig:Aannihilation}
\end{center}
\end{figure}

The pseudoscalar can be lighter or heavier than the singlino.
Assuming that DM is a thermal relic, the two DM components couple to each other via \Eq{W} and keep each other in thermal equilibrium.
The exact amount of the relic density is given by the contribution of several annihilation cross sections
since we have two DM particles with different annihilation modes.
The annihilation cross section into SM particles (fermions and vectors) is again suppressed by the smallness of the coupling $|\lambda_{sH}|$.
At leading order we have for the annihilation cross sections
\bea
&& \hspace{-1cm}  \sigma^\text{SM}_\psi v_\text{rel} \simeq \frac{v_\text{rel}^2}{4} \frac{g_s^2 \sin^2(2 \theta_{sH}) m_{\psi_S}^2  }{(4 m_{\psi_S}^2-m_h^2)^2+m_h^2 \Gamma_h^2(m_h)} \frac{\Gamma_h(2 m_{\psi_S}) }{2 m_{\psi_S}},\nonumber \\
&& \hspace{-1cm}  \sigma^\text{SM}_A v_\text{rel} \simeq \frac{ 2 \lambda_{sH}^2 v^2 }{(4 m_A^2-m_h^2)^2+m_h^2 \Gamma_h^2(m_h)} \frac{\Gamma_h(2 m_A) }{2 m_A},
\eea
where $v_\text{rel}$ is the relative DM velocity, $\Gamma_h(m)$ is the width of a SM Higgs of mass $m$.
In fact the dominant annihilation modes are the kinematically allowed processes among:
\be
 \psi_S  \psi_S \to s_i s_j, \quad  A A \to s_i s_j, \quad  A A \to \psi_S  \psi_S ,
\ee
where $s_{i,j}=s,h,A$ (see Figs. \ref{fig:psiannihilation} and \ref{fig:Aannihilation}).
If $CP$ is violated in the dark sector, perhaps by more complicated $W$ (or $V_\text{soft}$) involving complex parameters,
the pseudoscalar mixes with the Higgs and decays.

Because of the wide number of possible annihilation modes it is not straightforward to give an estimation of the total DM annihilation cross section.
We here consider a simplified configuration, leaving a more general analysis for a future work. We assume that $m_s < m_{\psi_S} < m_A$
and that $A$ is not stable (either because there is a small CP violating phase or $m_A>2 m_{\psi_S}$).
In this case the singlino will be the only DM candidate. Moreover since we still assume small mixing between $s$ and $h$, the relevant annihilation modes
are the ones in Fig. \ref{fig:psiannihilation} with $s_i=s_j=s_k=s$.
So we essentially reproduce the situation depicted in \cite{Farina:2013mla}.
Ignoring terms in $\lambda_{sH}$, the relevant annihilation cross section is
\be
 (\sigma v_\text{rel})_{\psi \psi \to ss} \simeq g_s^2 v_\text{rel}^2 \sqrt{1-\frac{m_s^2}{m_{\psi_S}^2}} \left(A_s^2 + A_{s(t+u)} + A_{t+u}^2 \right),
\label{sigmavss}
\ee
where $A_s$ is the contribution from the $s$-channel exchange of the scalar $s$ in Fig. \ref{fig:psiannihilation},
$A_{t+u}$ is the one from the $t(u)$-channel exchange of $\psi_S$ and $A_{s(t+u)}$ is the interference term
\bea
 A_s^2 &=& \frac{a_{s^3}^2}{128 \pi  \left(m_s^2-4 m_{\psi_S}^2\right)^2},\\
 A_{s(t+u)} &=& \frac{g_s m_{\psi_S} a_{s^3} \left(5 m_{\psi_S}^2-2 m_s^2\right)}{48 \pi  \left(4  m_{\psi_S}^2-m_s^2\right) \left(m_s^2-2 m_{\psi_S}^2\right)^2},\\
 A_{t+u}^2 &=& \frac{g_s^2 m_{\psi_S}^2 \left(9 m_{\psi_S}^4-8 m_{\psi_S}^2 m_s^2+2  m_s^4\right)}{24 \pi  \left(m_s^2-2 m_{\psi_S}^2\right)^4},
\eea
with $a_{s^3}=3! \, (\frac{a_\lambda}{\sqrt2} + 3\sqrt2 \mu_S \lambda_S + 9  v_s \lambda_S^2) $ the effective $s_R^3$ coupling.

\begin{figure}[t]
\begin{center}
\includegraphics[width=0.4\textwidth]{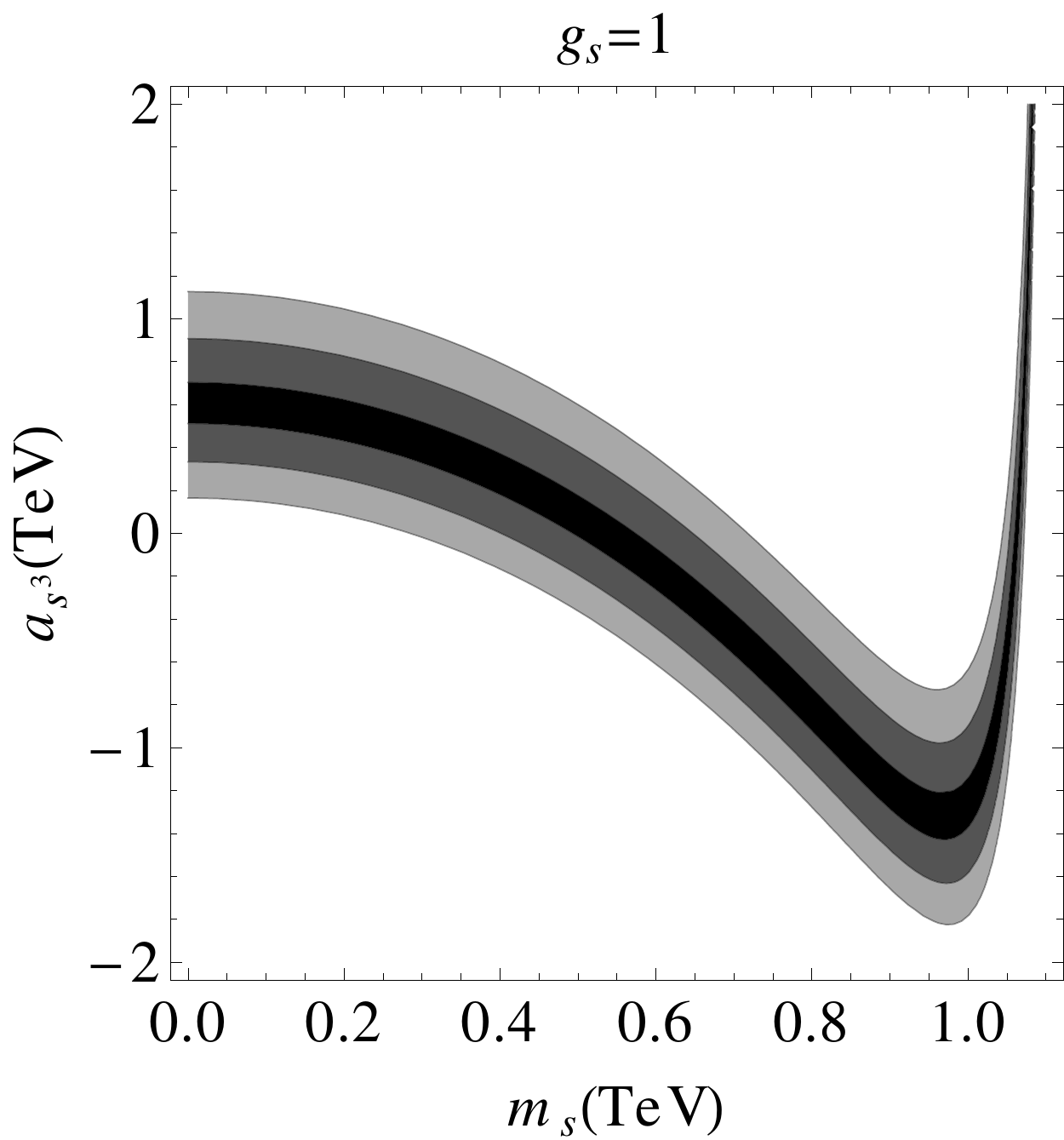}
\caption{Relic density estimation for $g_s=1$. The black region
corresponds to a relic density in the range $\Omega_c h^2 \pm
\sigma$, the darker gray to $\Omega_c h^2 \pm 3 \sigma$, the
lighter gray to $\Omega_c h^2 \pm 5 \sigma$, and the white region
is for relic densities out of any of the previous ranges.}
\label{fig:relicpos}
\end{center}
\end{figure}

\begin{figure}[t]
\begin{center}
\includegraphics[width=0.4\textwidth]{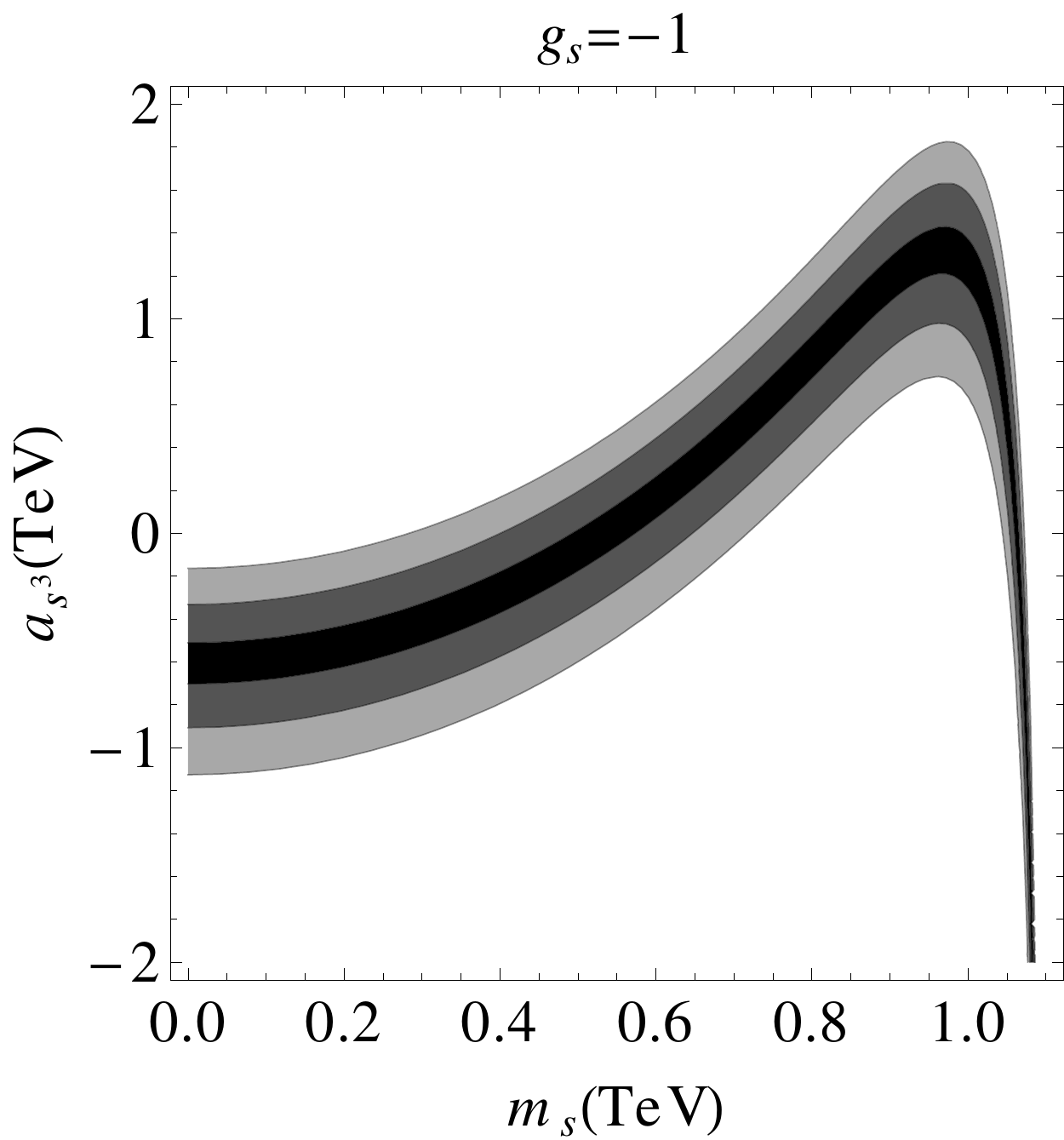}
\caption{Relic density estimation for $g_s=-1$. The black region
corresponds to a relic density in the range $\Omega_c h^2 \pm
\sigma$, the darker gray to $\Omega_c h^2 \pm 3 \sigma$, the
lighter gray to $\Omega_c h^2 \pm 5 \sigma$, and the white region
is for relic densities out of any of the previous ranges.}
\label{fig:relicneg}
\end{center}
\end{figure}

The Planck Collaboration \cite{Ade:2013lta} measured the cold DM relic density to be $\Omega_c h^2 \pm \sigma = 0.1199 \pm 0.0027$.
Assuming $|g_s|=1$ ($|\lambda_S| \simeq 0.24$) and $m_{\psi_S}=1.1$ TeV, we compute the corresponding relic density.
We present our results in Figs. \ref{fig:relicpos} and \ref{fig:relicneg} in the form of contour plots as functions of $m_s$ and $a_{s^3}$.
The black region corresponds to a relic density in the range $\Omega_c h^2 \pm \sigma$, the darker gray to $\Omega_c h^2 \pm 3 \sigma$,
the lighter gray to $\Omega_c h^2 \pm 5 \sigma$, and the white region is for relic densities out of any of the previous ranges.
The cross section (\ref{sigmavss}) depends on the relative sign between $g_s$ and $a_{s^3}$.
In fact Fig. \ref{fig:relicneg} can be obtained from Fig. \ref{fig:relicpos} by applying a reflection along the $a_{s^3}=0$ axis.
We have a relevant region in which we have agreement with experimental data and we can appreciate the relevant role of the parameter $a_{s^3}$,
in contrast to the model described in \cite{Farina:2013mla}, where such a term is negligible.
We can see that low $m_s$ ($\lesssim 300$ GeV) is favored when $g_s$ and $a_{s^3}$ have the same sign,
while high $m_s$ ($\gtrsim 700$ GeV) is favored when $g_s$ and $a_{s^3}$ have opposite sign.
This last situation is most consistent with our assumption of small mixing angle $\theta_{sH}$.
Finally we stress that the natural configuration $m_{\psi_S}$, $m_s$, $|a_{s^3}| \sim 1$ TeV is allowed.
\begin{figure}[t]
\begin{center}
\includegraphics[width=0.5\textwidth]{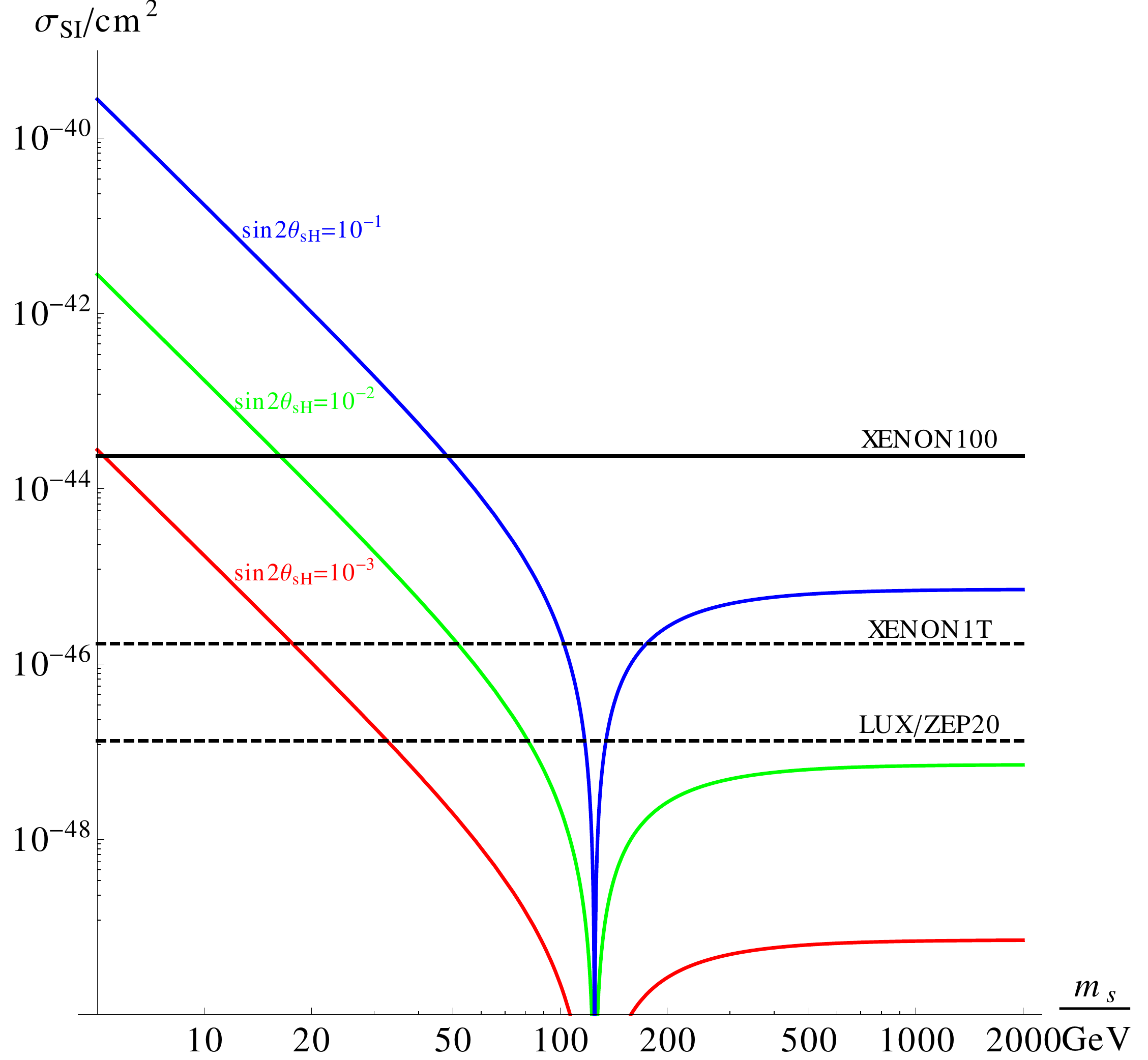}
\caption{Singlino direct detection cross section for $|g_s|=1$ and $m_{\psi_S}=1.1$ TeV  in function of $m_s$  for $\sin(2\theta_{sH})=10^{-1}$ (blue),
         $\sin(2\theta_{sH})=10^{-2}$ (green) $\sin(2\theta_{sH})=10^{-3}$ (red).
         The black continuous line represents XENON100 bound for 2012, while the two black dashed lines stands for XENON1T and LUX/ZEP20 projections.}
\label{fig:SIplot}
\end{center}
\end{figure}

To conclude we stress again that the singlino DM is almost undetectable in the recent direct detection experiments,
but there are some chances in the future experiments. In order to show that, we plot in Fig. \ref{fig:SIplot} the cross section (\ref{sigmaSIpsi})
as a function of $m_s$, for three reference values of $\sin\theta_{sH}$, assuming again $|g_s|=1$ and $m_{\psi_S}=1.1$ TeV.
The blue line represents $\sin(2\theta_{sH})=10^{-1}$, the green one $\sin(2\theta_{sH})=10^{-2}$ and the red one $\sin(2\theta_{sH})=10^{-3}$.
The black continuous line represents the bound coming from XENON100 data \cite{Aprile:2012nq}, while the two black dashed lines stands for XENON1T \cite{Aprile:2012zx} and LUX/ZEP20 \cite{LUXZEP} projections\footnote{To produce the curves,
we used the online tool at http://dendera.berkeley.edu/plotter/entryform.html}.
Such experimental lines depend on the DM mass. In our case DM mass is fixed to be $m_{\psi_S}=1.1$ TeV, so in our plot they will look as horizontal lines.
We can see that with a tiny mixing angle ($\sin(2\theta_{sH})\lesssim 10^{-2}$),
the direct detection cross section is always (but for $m_s \lesssim$ 20 GeV) not excluded by the XENON100 bound.
Moreover there are chances to detect it in future experiments if the scalar singlet is relatively light
($m_s \lesssim$ 50 GeV) even if the mixing is small.
However the future detection for $m_s \gtrsim$ 50 GeV is possible only if the mixing is not anymore suppressed.

Finally we remind that our results are considering only a small region ($|g_s|=1$ and $m_{\psi_S}=1.1$ TeV) of the full parameters space and that
we will present a more detailed study in a future work where all the possible configurations of our scenario will be taken into account.


\section{Discussion and conclusions}

In this paper we have proposed that Dark Supersymmetry is responsible for EW symmetry breaking in the SM and for the DM of the Universe.
Such a phenomenological setup, consistent with the absence of new physics signals in the LHC and flavor physics data, consists of two sectors: The classically conformal
but non-supersymmetric SM sector and the dark sector that is supersymmetric. The first may originate from a recently proposed
supersymmetry breaking scenario in which all superpartners are confined to a separate brane in extra dimensions and are completely absent in our world.
In our scenario we propose that the mass scale in the dark sector, which is stable against radiative corrections, is transferred to the visible
sector via the Higgs portal. The resulting theory is the SM, in which the Higgs boson has a permille level mixing with a singlet scalar that is the
messenger from the dark sector, plus the DM of the Universe. We have shown that for TeV scale Dark Supersymmetry the observed
DM relic density is naturally generated in this scenario despite of the strongly suppressed DM coupling to baryonic matter. The latter
occurs only through Higgs mixing and explains the negative results of DM direct detection experiments.

The phenomenology of this class of models is quite generic; any consistent dynamical mechanism that generates a stable mass scale in the dark sector
results in a similar theory from the SM sector point of view. Dark Supersymmetry is just one possibility, the other known proposals are Dark Technicolor and
Dark Coleman-Weinberg. Although the physics in dark sectors is very different in all those models, they all couple to the SM only via the Higgs portal.
Therefore testing this class of models directly is extremely challenging. Perhaps the best chance is to perform precision studies of the Higgs boson
couplings at the linear collider, and to compare direct measurements of Higgs couplings with the measured particle masses in order to detect any deviation from the SM predictions.

We would like to conclude with discussing possible extensions of the most minimal scenario presented in this work.
The dynamics and particle content of the dark sectors are expected to  be rather complicated.
It might also be possible that the Higgs portal couplings to the dark sector are not of the minimal form presented in \Eq{portal}.
Other scalar messenger fields $S_i$ to the dark sector may exist that may carry SM quantum numbers.
If the scalar messengers $S_i$ couple to other SM particles in addition to the Higgs boson, those non-minimal models have
much brighter prospects to be tested by the LHC and by DM direct detection experiments.
The non-minimal portals may, for example, be motivated by explaining other SM features with dark sector dynamics,
such as flavor physics, that have no explanation in the context of the SM. We therefore conclude that the proposed
scenario has a rich phenomenology and that further studies along those lines are needed.

\vspace{0.5cm}

\mysection{Acknowledgement}
We thank A. Strumia and K. Kannike for discussions.
This work was supported by the ESF grants 8499, 8943, MTT8, MTT59, MTT60, MJD140, MJD435, MJD298,
by the recurrent financing SF0690030s09 project and by the European Union through the European Regional Development Fund.

\end{document}